\documentclass[10pt, twocolumn,secnumarabic,amssymb, nobibnotes, aps,pra longbibliography,superscriptaddress]{revtex4-2}
\usepackage[utf8]{inputenc}
\usepackage[T1]{fontenc}
\usepackage{bm}
\usepackage{color}
\usepackage{subfigure}
\usepackage[pdftex]{graphicx}

\newcommand{\angstrom}{\textup{\AA}}
\newcommand{\electronvolt}{\textup{eV}}
\newcommand{\volt}{\textup{V}}
\newcommand{\Pt}{\textup{Pt}}
\newcommand{\Au}{\textup{Au}}
\newcommand{\Co}{\textup{Co}}
\newcommand{\C}{\textup{C}}

\newcommand{\mili}{\textup{m}}
\newcommand{\hBN}{\textup{h-BN}}

\newcommand{\Gr}{\textup{Gr}}
\newcommand{\sysau}{\Co/\Au/\Gr}
\newcommand{\syspt}{\Co/\Pt/\Gr}
\newcommand{\Ni}{\textup{Ni}}

\begin{document}
\title{Proximity-induced magnetization in graphene: Towards efficient spin gating}
\author{Mihovil Bosnar}
\email{mbosnar@irb.hr}
\affiliation{Ru\dj{}er Bo\v{s}kovi\'{c} Institute, 10000 Zagreb, Croatia}
\author{Ivor Lon\v{c}ari\'{c}}
\email{ivor.loncaric@gmail.com}
\affiliation{Ru\dj{}er Bo\v{s}kovi\'{c} Institute, 10000 Zagreb, Croatia}
\author{P. Lazi\'{c}}
\affiliation{Research Computing Support Services group, University of Missouri, Columbia, MO, 65211-7010, USA}
\affiliation{Ru\dj{}er Bo\v{s}kovi\'{c} Institute, 10000 Zagreb, Croatia}
\author{K. D. Belashchenko}
\affiliation{University of Nebraska-Lincoln and Nebraska Center for Materials and Nanoscience, Lincoln, NE 68588-0299, USA}
\author{Igor \v{Z}uti\'{c}}
\affiliation{University at Buffalo, State University of New York, Buffalo, NY 14260-1500, USA}

\begin{abstract}
Gate-tunable spin-dependent properties could be induced in graphene at room temperature through magnetic proximity effect by placing it in contact with a metallic ferromagnet. Because strong chemical bonding with the metallic substrate makes gating ineffective, an intervening passivation layer is needed. Previously considered passivation layers result in a large shift of the Dirac point away from the Fermi level, so that unrealistically large gate fields are required to tune the spin polarization in graphene. We show that a monolayer of Au or Pt used as the passivation layer between Co and graphene brings the Dirac point closer to the Fermi level. In the $\syspt$ system the proximity-induced spin polarization in graphene and its gate control are strongly enhanced by the presence of a surface band near the Fermi level. Furthermore, the shift of the Dirac point could be eliminated entirely by selecting submonolayer coverage in the passivation layer.
Our findings open a path towards experimental realization of an optimized two-dimensional system with gate-tunable spin-dependent properties.
\end{abstract}

\maketitle

\section{Introduction}
With its high mobility and low spin-orbit coupling, graphene (Gr) is expected to be a particularly suitable material for spin transport and spintronics~\cite{Han2014:NN,Lin2019:NE,Avsar2020:RMP}. From the first demonstration of spin injection in graphene~\cite{Tombros2007:N}, there has been significant progress in extending the characteristic time and length scales over which spin information can be sustained~\cite{Drogeler2014:NL,Zomer2012:PRB}. Graphene-based spin-logic gates have been demonstrated at room temperature~\cite{Wen2016:PRA}, supporting proposals for specialized applications that could outperform CMOS-based counterparts~\cite{Dery2012:IEEETED}. 

Despite its attractive intrinsic properties, there is a strong interest in introducing superconductivity, magnetism, a sizeable energy gap, topological properties, or strong spin-orbit coupling into graphene. Chemical doping or functionalization tend to introduce unwanted disorder and significantly reduce the mobility of charge carriers in graphene~\cite{Zutic2019:MT}. However, the atomic thickness of graphene and other two-dimensional (2D) materials offers an alternative way to modify their properties through short-range proximity effect~\cite{Zutic2019:MT} from an adjacent layer that already has the desired properties~\cite{Heersche2007:N,Wei2016:NM,Gmitra2016:PRB,Avsar2014:NC,Rossi2020:AP,Lee2016:PRB,PhysRevMaterials.2.054002}.

\begin{figure}[b!]
\centering
\includegraphics[scale=1.0]{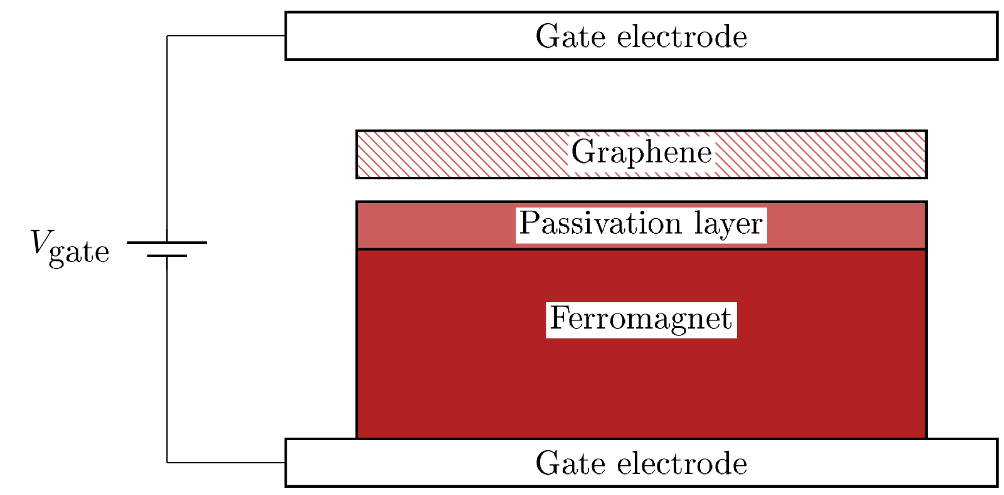}
\caption{Schematic view of the proposed system.}
\label{fig:scheme}
\end{figure}

Interest in magnetic proximity effects in 2D systems is exceedingly broad, as they are considered for implementing magnetic skyrmions~\cite{Yang2018:NM}, exotic properties of topological insulators~\cite{Qi2011:RMP}, and Majorana bound states for topological quantum computing~\cite{Sau2010:PRL,Fatin2016:PRL,Zhou2019:PRB,Desjardins2019:NM}. In this paper we focus on magnetic-proximity effects in Co/Gr-based hetrostructures, of the type shown in Fig.~\ref{fig:scheme}, that could enhance graphene for spintronic applications.

Many graphene-based spintronic devices operate by switching the magnetization in ferromagnetic contacts~\cite{Han2014:NN,Lin2019:NE,Avsar2020:RMP,Lazic2014:PRB,Raes2017:PRB,Ringer2018:APL}. As an alternative to the use of an external magnetic field, it is attractive to take advantage of the electrically-tunable magnetic proximity effect in graphene, which is potentially faster and more energy-efficient~\cite{Lazic2016:PRB,Zhao2019:2DM}.
To characterize this effect, we use the proximity-induced spin polarization $P(E)$ in graphene, which is defined as
\begin{equation}
    P(E)=\frac{N_{\uparrow}(E)-N_{\downarrow}(E)}{N_{\uparrow}(E)+N_{\downarrow}(E)},
    \label{polarization}
\end{equation}
where $N_s(E)$ is the projected density of states (PDOS) for spin projection $s$ averaged over the $\C$ atoms.
Both magnitude and sign of $P(E)$ near the Fermi level $E_F$ can be controlled by electric gating~\cite{Lazic2016:PRB}.

It is usually expected that tunable magnetic proximity effects require a magnetic insulator to avoid shorting the circuit through the metallic ferromagnet~\cite{Yang2013:PRL,Wang2015:PRL,Swartz2012:ACSN,Behera2019:PCCP,Takiguchi2019:NP,Cortes2019:PRL}. Indeed, graphene forms strong chemical bonds with Co and Ni~\cite{Lazic2016:PRB,Zollner2016:PRB,Giovannetti2008:PRL}, which essentially turns it into a metallic continuation of the ferromagnet with a large DOS at the Fermi level.
As a result, it is practically impossible to change $P(E_F)$ in such chemisorbed graphene by electric gating~\cite{Lazic2016:PRB}. 

Therefore, the idea to select heterostructures with a van-der-Waals (vdW) bonded layer of graphene was introduced~\cite{Lazic2016:PRB}. A common metallic ferromagnet with high Curie temperature could then be considered for tunable magnetic proximity effects. In particular, surface passivation through an addition of another graphene layer~\cite{Lazic2016:PRB} or a layer of h-BN~\cite{Lazic2016:PRB,Zollner2016:PRB} between graphene and Co, as shown in Fig.~\ref{fig:scheme}, was studied. In both cases passivation results in weak vdW bonding of the top layer of graphene to the underlying structure~\cite{Lazic2016:PRB}.

Weak vdW bonding tends to preserve major features of the electronic structure, such as the Dirac cone in graphene~\cite{Lazic2016:PRB,Giovannetti2008:PRL,Asshoff2017:2DM}, but the bound system still has important differences compared to its standalone components. In systems studied previously, the most relevant features are ($n$-type) doping~\cite{Lazic2016:PRB,Giovannetti2008:PRL} and lifting of the spin degeneracy in graphene due to the proximity effect~\cite{Lazic2016:PRB,Zollner2016:PRB}. This proximity effect has been utilized in transport experiments~\cite{Asshoff2017:2DM} aiming to use graphene not as a spin filter (as in the tunneling geometry~\cite{Karpan2007:PRL}) but as a \emph{source} of spin-polarized carriers.

Graphene and the passivated surface can be thought of as two plates of a capacitor. Gate voltage $V_{gate}$ induces an electrostatic potential difference and charge transfer between these plates, which can change $P(E_F)$ in graphene and sometimes even its sign~\cite{Lazic2016:PRB}. The potential difference depends inversely on the DOS at the Fermi level in graphene~\cite{Lazic2016:PRB}. It is, therefore, expected that $P(E_F)$ should be more sensitive to gating if the Fermi level is close to the Dirac point.

\begin{figure*}[t!]
\centering
\begin{subfigure}
\centering
\includegraphics[scale=1.0]{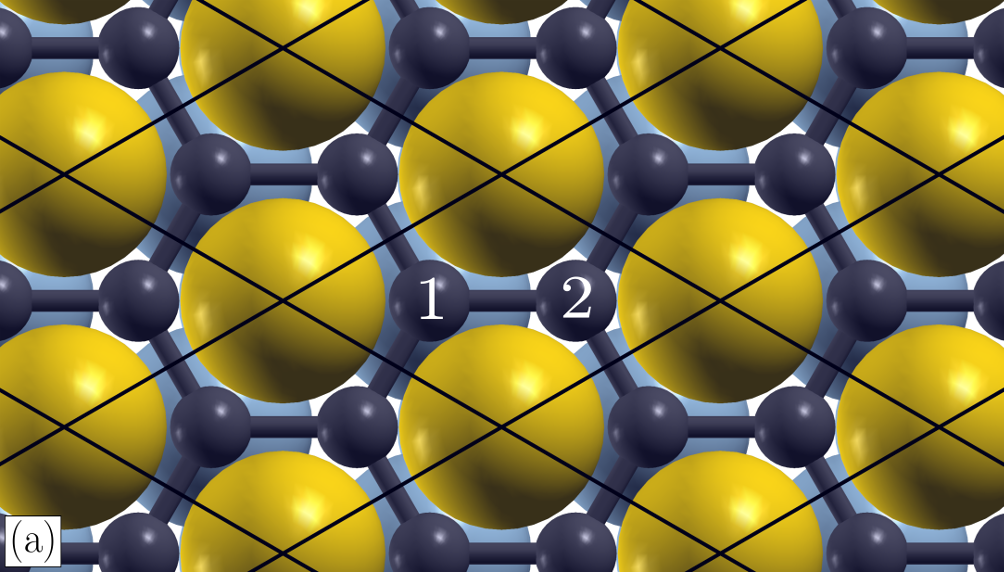}
\end{subfigure}
\qquad
\begin{subfigure}
\centering
\includegraphics[scale=1.00]{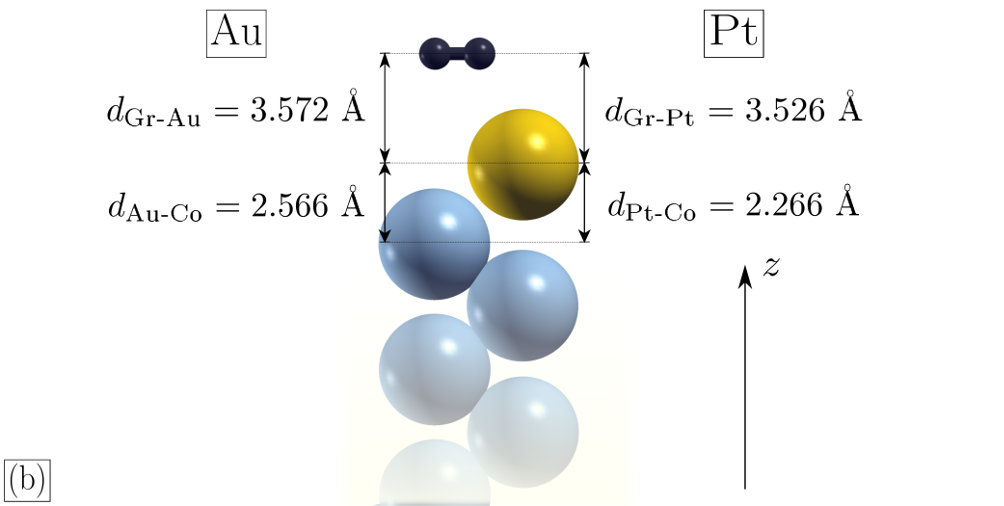}
\end{subfigure}
\\
\vspace{0.25cm}
\begin{subfigure}
\centering
\includegraphics[scale=1.0]{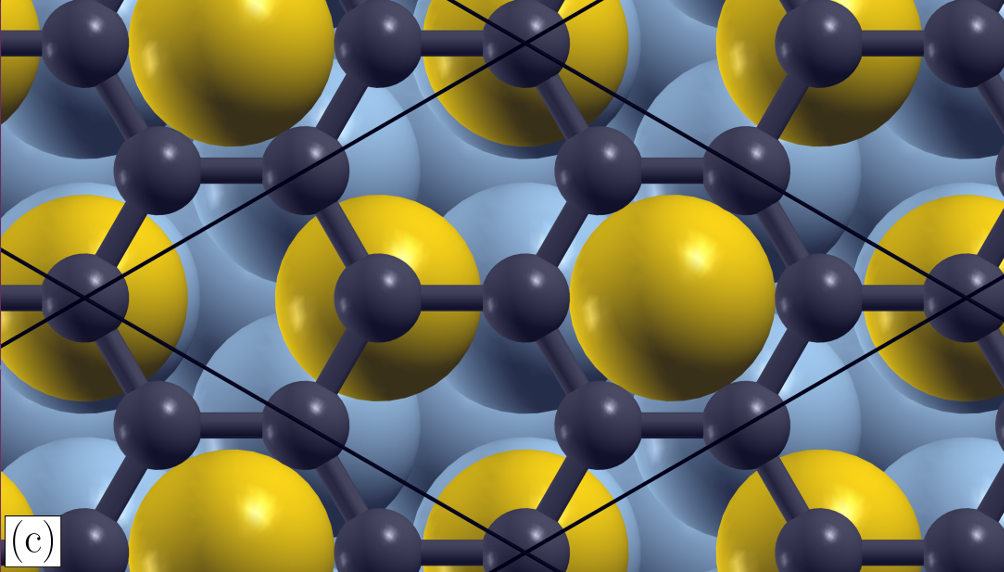}
\end{subfigure}
\qquad
\begin{subfigure}
\centering
\includegraphics[scale=1.00]{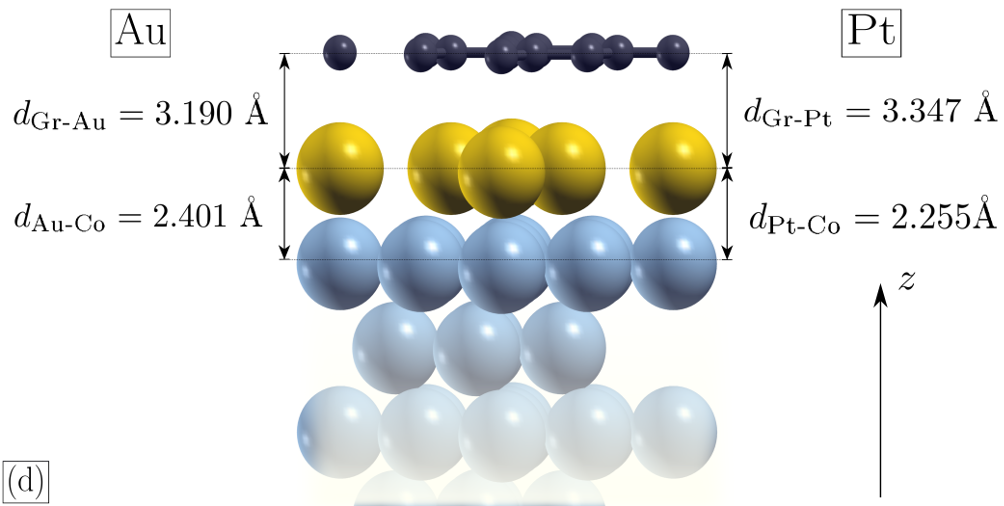}
\end{subfigure}
\caption{Top (a) [(c)] and side (b) [(d)] view of the optimal energy structure with $100\%$ [$75\%$] $\Co$ surface coverage by $\Pt$ or $\Au$. $\C$ atoms are shown in dark blue, noble metal in yellow, and $\Co$ in light blue. Two inequivalent $\C$ atoms in case of $100\%$ coverage are marked by $1$ and $2$. The listed average interlayer distances correspond to equilibrium without the applied field. This figure was made using XCrySDen software~\cite{xcrysden}.}
\label{system}
\end{figure*}

With h-BN and graphene passivation layers on the surface of Co, physisorbed graphene is $n$-doped with its Dirac point shifted approximately $0.5\ \electronvolt$ below the Fermi level and a relatively large DOS at the Fermi level~\cite{Lazic2016:PRB}. First-principles calculations have shown that significant tuning of $P(E)$ in this system requires large electric fields~\cite{Lazic2016:PRB} that are achievable only by ionic-liquid gating~\cite{Hebard1987:IEEETM,Mannhart1993:APL}. This conclusion was corroborated by experiments on lateral Co/h-BN/Gr-based spin valves~\cite{Xu2018:NC}, where electrostatic gating failed to produce tunable magnetic proximity effects in 2D ferromagnetic contacts in the planar geometry similar to Fig.~\ref{fig:scheme}. Instead, to reduce detrimental $n$-doping, one-dimensional edge contacts were used to achieve the ability to switch the proximity-induced spin polarization in graphene by gating~\cite{Xu2018:NC}.

In this paper we explore a different path to realizing tunable spin polarization in 2D geometry by using a monolayer of noble-metal atoms like Au or Pt as the passivation layer between the Co surface and graphene. Furthermore, we show that additional flexibility can be achieved by using a passivation layer with a variable partial coverage. In fact, a study of graphene doping on $\Au$-passivated $\Ni$ surface found that reducing the coverage of $\Ni$ by $\Au$ atoms reduces the doping level in graphene~\cite{Kang2010:PRB}. Our first-principles calculations based on these considerations provide useful guidance for realizing proximity-induced spin polarization and spin gating in graphene.

While doping of graphene is not eliminated in any of the structures studied here, all of them have more than an order of magnitude larger $P(E_F)$ at electric fields under $0.1\ \volt/\angstrom$ compared to Co/h-BN/Gr heterostructures, reaching $15 \%$ with Pt-passivated $\Co$. Furthermore, doping in $100\%$ covered structures is $p$-type, in contrast to $n$-type in the $\Co/\hBN/\Gr$ structure, while the reduction in coverage drastically shifts the doping level towards $n$-type for both $\Au$ and $\Pt$-passivated $\Co$. This result suggests a way to design the desired structure where doping is entirely cancelled.

\section{Computational details}

Proximity-induced spin polarization in graphene was studied using density functional theory (DFT) as implemented in a real-space code GPAW~\cite{gpaw1,gpaw2,ase}. The real-space approach avoids the use of periodic boundary conditions in the non-periodic direction, preventing spurious tunneling ~\cite{Lazic2016:PRB} and obviating the need for dipole interaction corrections~\cite{Hakan2013:JPCM}. 

We used projector-augmented-wave (PAW)~\cite{paw1, paw2} PBE~\cite{pbe} setups from the GPAW package and the semilocal vdW functional vdw-df-cx~\cite{cx1, cx2,PhysRevMaterials.3.063602} from the libvdwxc library~\cite{libvdwxc} for exchange and correlation. The cell was sampled by a grid of 0.133 (0.129) $\angstrom$ spacing in the planes parallel (perpendicular) to the heterostructure layers.

Input structures were constructed in QuantumWise Virtual NanoLab~\cite{vnl}. For the fully covered hexagonal close-packed $\Co\ (0001)$ surface we used a slab with $7$ layers of $\Co$ atoms. The passivating $\Au$ or $\Pt$ monolayers were added in an extension of this slab, fully covering the surface and forming a $1\times1$ supercell. Since the experimental lattice constants of graphene ($2.461\ \angstrom$) and $\Co$ $(0001)$ surface ($2.507\ \angstrom$) make a small lattice mismatch of $1.8\%$, we used a $1\times1$ supercell with the lattice parameter of graphene. All supercells contained $10\ \angstrom$ of vacuum on both sides of the heterostructure. 

To find the optimal lateral position of graphene, we relaxed the system starting from three configurations: $\Au$ or $\Pt$ atom located under the hollow site of graphene, under a C-C bond, and under a C atom. Relaxations were performed using the quasi-Newton algorithm, as implemented in the ASE package~\cite{ase}, with GPAW as the DFT calculator, until the force on each atom was smaller than $0.05\ \electronvolt/\angstrom$. We fixed the bottom two layers of $\Co$ and used a $15\times15\times1$ Monkhorst-Pack sampling of the Brillouin zone with a Fermi-Dirac smearing of $200\ \mili\electronvolt$.

The total energy for each relaxed structure was calculated with a dense $63\times63\times1$ k-point mesh and Fermi-Dirac smearing of $10\ \mili\electronvolt$. The optimal configuration for both passivation layers is with the $\Au$ or $\Pt$ atoms located under the hollow site of graphene. For $\sysau$ ($\syspt$), the bond structure is  $21.9$ ($96.7$) $\mili\electronvolt$ higher in energy, and the on-top structure $56.1$ ($24.8$) $\mili\electronvolt$ higher than the hollow-site structure. The optimal structures are shown in Figs.~\ref{system}(a) and \ref{system}(b). These structures were relaxed further with a homogeneous electric field of $E_z=\pm0.01$, $\pm0.05$, $\pm0.1$, and $\pm0.2\ \volt/\angstrom$, applied perpendicular to the surface.

As a case study of a system with lower coverage in the passivation layer, we used a system with a $75\%$ covered $\Co$ surface shown in Figs.~\ref{system}(c) and \ref{system}(d). To obtain the $75\%$ coverage, $3$ atoms of $\Au$ or $\Pt$ were placed on the $2\times2$ supercell of $\Co\ (0001)$. The starting positions were taken from Ref.~\cite{Kang2010:PRB} for the Ni(111)/Au/Gr system. Instead of $7$ $\Co$ layers used for the full coverage, here we used 5 $\Co$ layers to keep the calculation manageable; the system was relaxed with the same procedure. 

\begin{figure*}
\centering
  \begin{subfigure}
    \centering
    \includegraphics[scale=1.0]{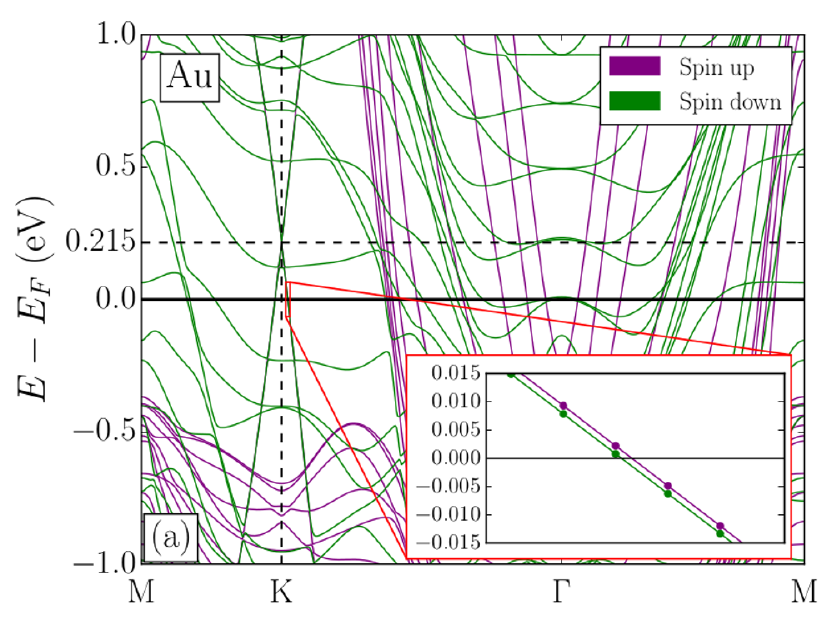}
   \end{subfigure}
  \qquad
   \begin{subfigure}
    \centering
    \includegraphics[scale=1.0]{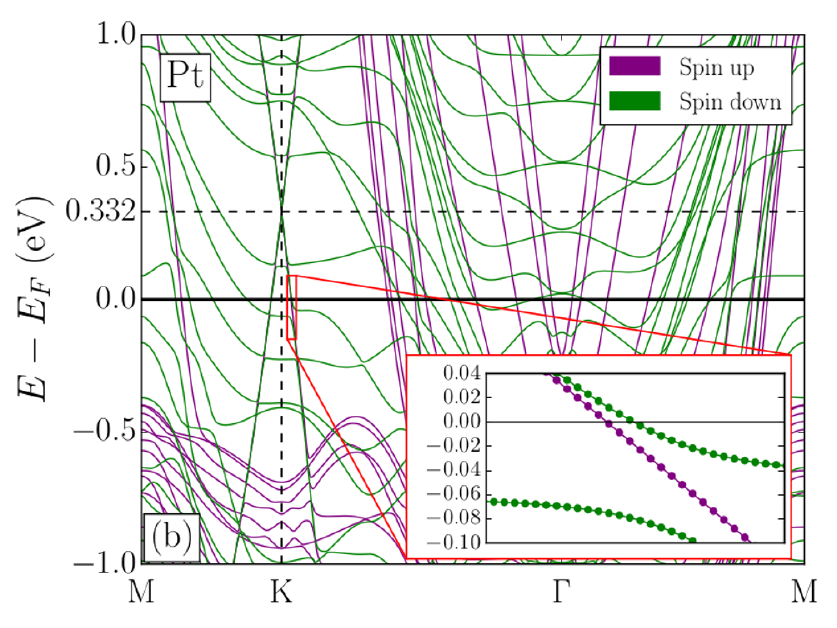}
   \end{subfigure}  
   \caption{The band structure of (a) $\sysau$ and (b) $\syspt$ for full coverage in the passivation layer. Insets: enlarged portion of the band structure near the Fermi level. 
   }
   \label{bandstructure}
\end{figure*}

We used finer Brilloiun zone sampling and Fermi-Dirac smearing of $10\ \mili\electronvolt$ for the precise calculation of the magnetic moments. For systems with $100\%$ ($75\%$) coverage we used a $111\times111\times1$ ($32\times32\times1$)  k-point mesh,
which was also used to calculate the spin-resolved PDOS where $200\ \mili\electronvolt$ Gaussian broadening was chosen.

The atomic magnetic moments are estimated as the integrals of the spin density $\mathcal{S}$, defined as the difference of electronic densities $\rho$ in the two spin channels
\begin{equation}
    \mathcal{S}(\textbf{r})=\rho_{\uparrow}(\textbf{r})-\rho_{ \downarrow}(\textbf{r}),
    \label{spin_density}
\end{equation}
over the augmentation spheres.

\section{Results and discussion}
\subsection{Zero gate voltage}

The band structures of the $\sysau$ and $\syspt$ systems with  $100\%$ coverage in the passivation layer, at zero applied field, are shown in Fig.~\ref{bandstructure}. In both cases the Dirac cone of graphene is preserved but shifted to higher energies by $0.33\ \electronvolt$ and $0.22\ \electronvolt$ for $\Pt$ and $\Au$, respectively. Along with the distances of about $3.5\ \angstrom$ between graphene and the $\Pt$ or $\Au$ layer (see Fig.~\ref{system}), these results suggest vdW bonding between graphene and the metallic surface. Graphene here is $p$-doped to a lesser extent compared to $n$-doped graphene in the system with a h-BN passivation layer~\cite{Lazic2016:PRB}. 

\begin{figure*}
\centering
  \begin{subfigure}
    \centering
    \includegraphics[scale=1.0]{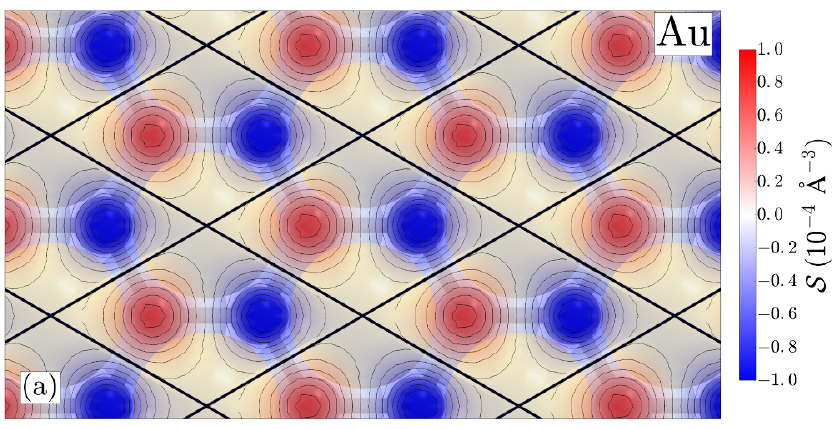}
   \end{subfigure}
   \begin{subfigure}
    \centering
    \includegraphics[scale=1.0]{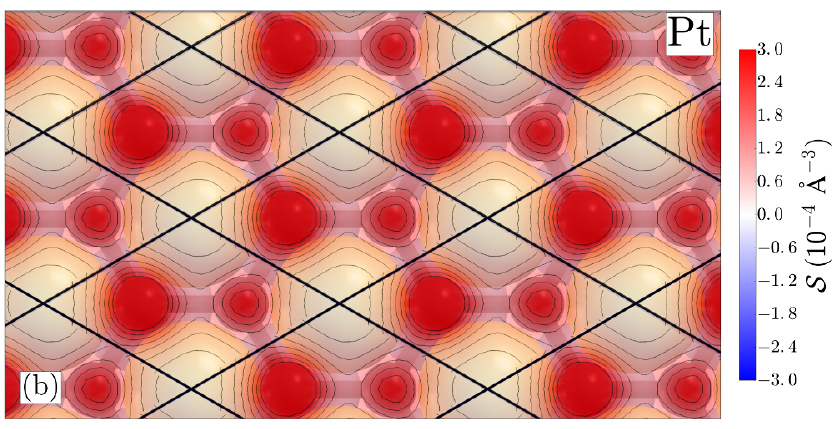}
   \end{subfigure}  
   \caption{Spin density $\mathcal{S}(\textbf{r})$ [see Eq.~(\ref{spin_density})] in (a) $\sysau$ and (b) $\syspt$ in a plane $0.33\ \angstrom$ above the graphene layer for 100\% coverage in the passivation layer. Red (blue) areas: same (opposite) sign of the spin density as in $\Co$. The colorbar scales are different.
   Figure was made using the XCrySDen software~\cite{xcrysden}.}
   \label{spin_density_fig}
\end{figure*}
\begin{figure*}
    \centering
    \begin{subfigure}
      \centering
      \includegraphics[scale=0.4]{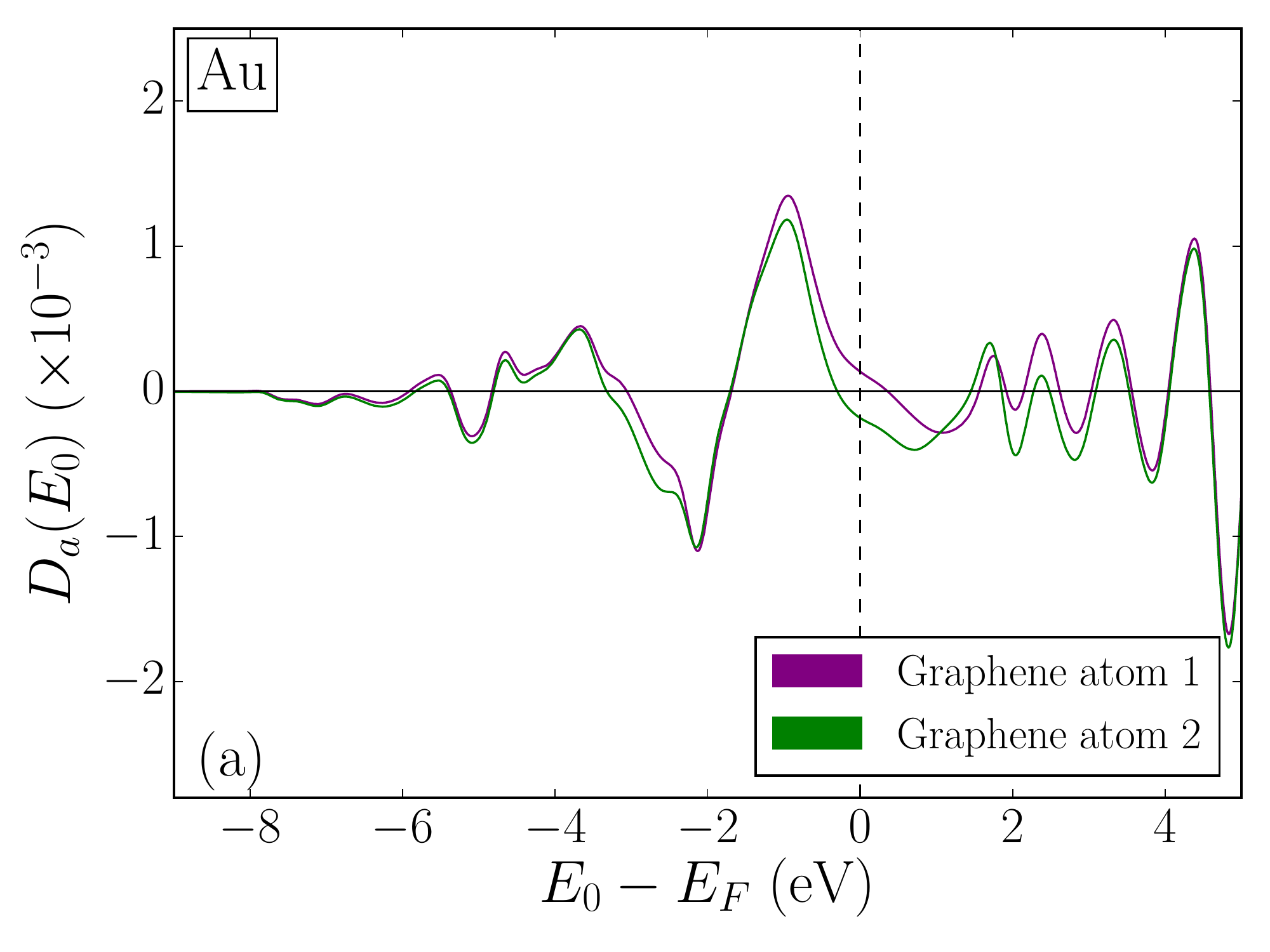}
    \end{subfigure}
    \qquad
    \begin{subfigure}
      \centering
      \includegraphics[scale=0.4]{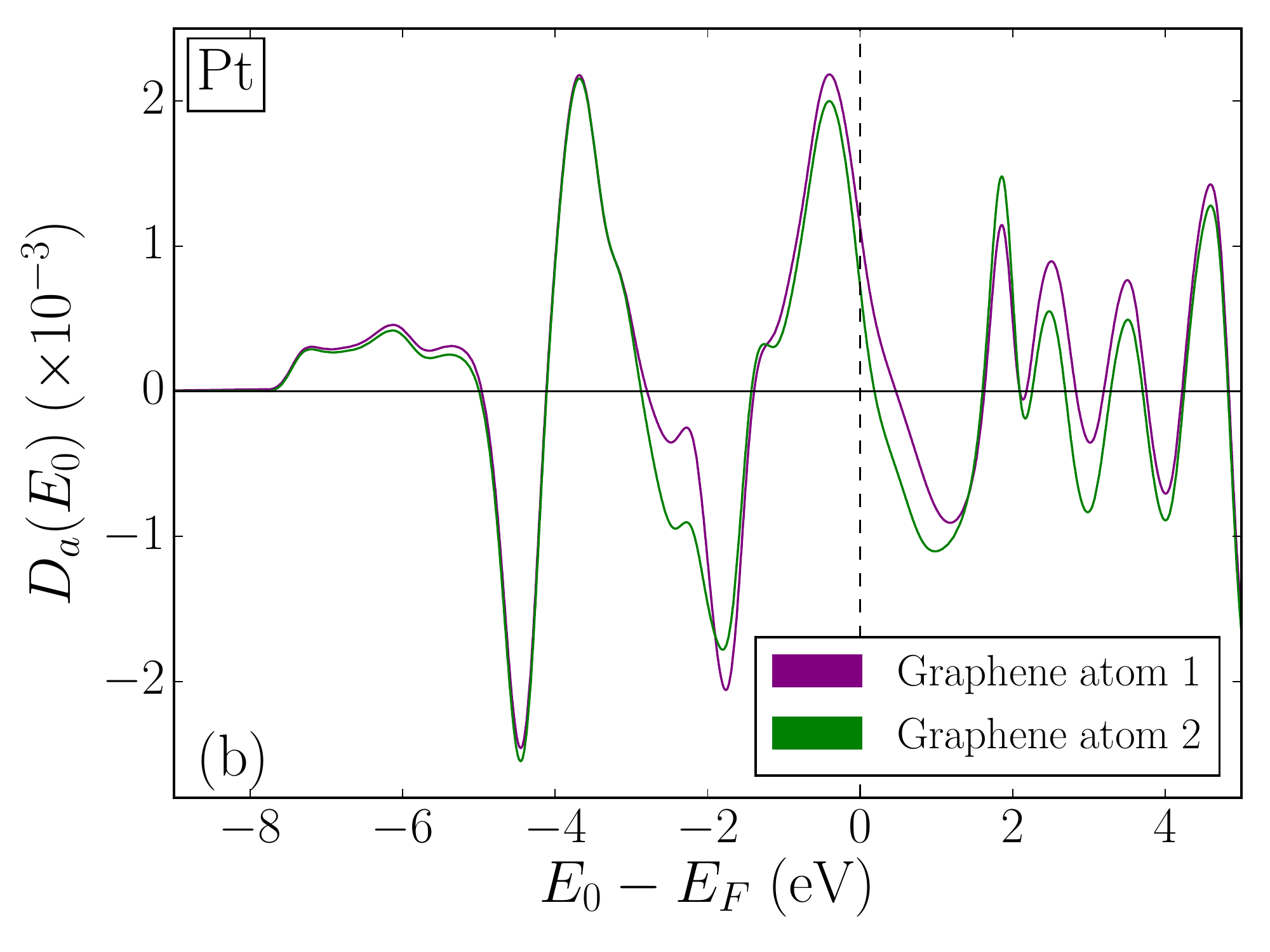}
    \end{subfigure}
    \caption{$D_a(E_0)$, defined in Eq.~(\ref{integral}), as a function of $E_0$, for 100\% coverage in the passivation layer. The graphene atoms are labeled as shown in Fig.~\ref{system}(a).}
    \label{integrated_PDOS}
\end{figure*}

A closer look at the band structure near the Fermi level shown in the insets in Figs.~\ref{bandstructure}(a) and \ref{bandstructure}(b) reveals that spin splitting is induced in graphene in both systems. In $\sysau$ the splitting is approximately constant near the Fermi level and we estimate it to be about $1.05\ \mili\electronvolt$. This splitting represents interlayer exchange~\cite{Zollner2016:PRB}. In contrast, in $\syspt$ the splitting varies sharply near the Fermi level due to the hybridization of the graphene and flat metal bands and the ensuing avoided band crossings. We note that avoided crossings are also present in $\sysau$, and exchange splitting is present in $\syspt$, but their effect on the respective band structures appears to be smaller.

The magnetic proximity effect induces a finite spin density $\mathcal{S}$ in graphene. Figure~\ref{spin_density_fig} shows a planar slice of $\mathcal{S(\mathbf{r})}$ at a distance of $0.33\ \angstrom$ above the carbon nuclei. Clearly, the spin density is inhomogeneous in both systems, reflecting the inhomogeneity of the surface. Furthermore, the spin density in $\sysau$ has different signs on the two inequivalent carbon atoms.

To gain insight into how $\mathcal{S}$ is formed from the band contributions, one can integrate the difference in PDOS $N_{a,s}(E)$ on a $\C$ atom of type $a$ up to some energy $E_0$
\begin{equation}
   D_a(E_0)=\int_{-\infty}^{E_0}\left[N_{a,\uparrow}(E)-N_{a,\downarrow}(E)\right] dE.
   \label{integral}
 \end{equation}
The resulting quantities $D_a(E_0)$ are shown in Fig.~\ref{integrated_PDOS} for the two inequivalent C atoms; the values $D_a(E_F)$ at the Fermi level are consistent with Fig.~\ref{spin_density_fig}. We see that $D_a(E_0)$ are complicated functions that change sign multiple times in both systems. This means that the spin density stems not from homogeneous exchange splitting, but from the hybridization of graphene bands with the $d$ bands of the metallic surface.

\begin{figure*}
\centering
  \begin{subfigure}
    \centering
    \includegraphics[scale=0.21]{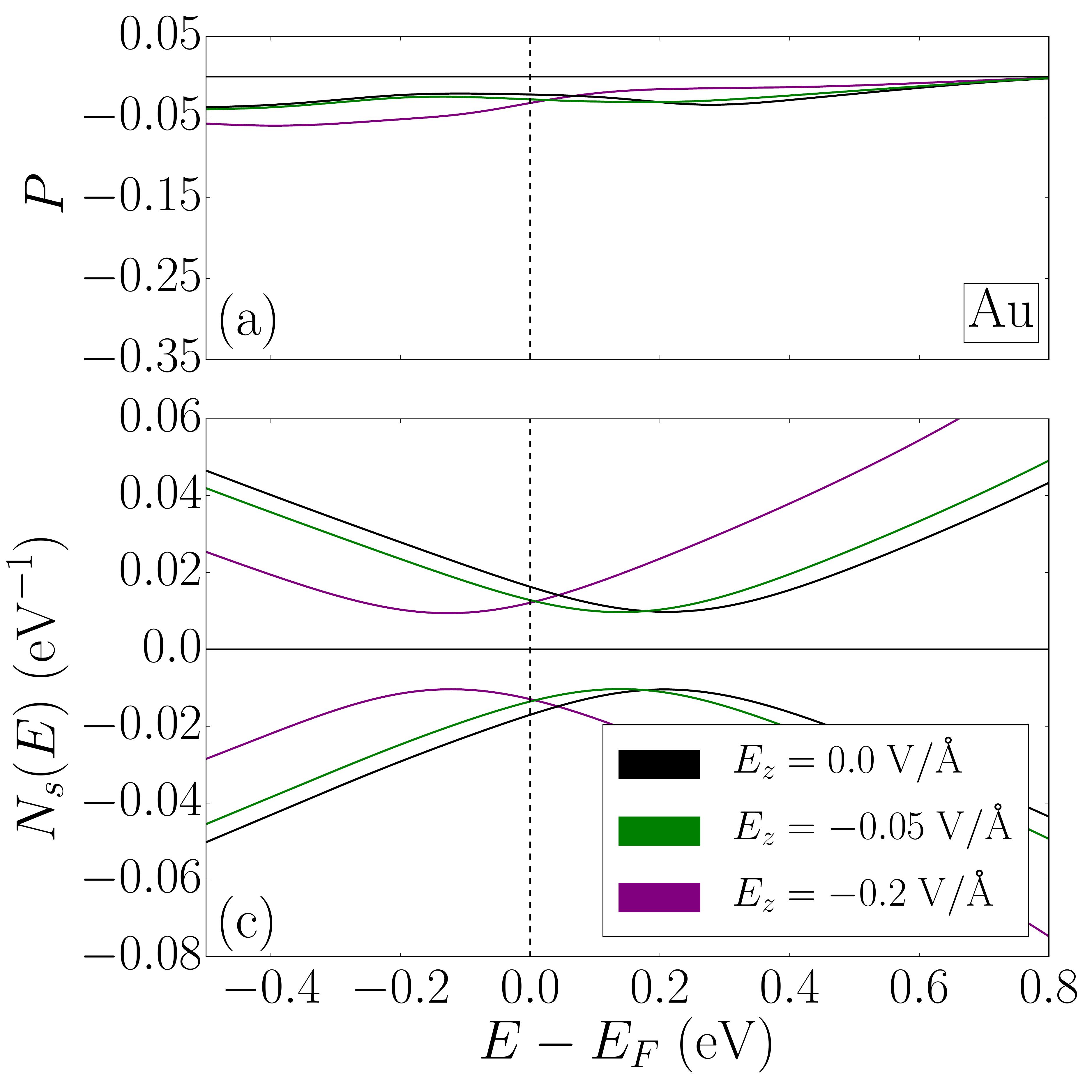}
   \end{subfigure}
\qquad
   \begin{subfigure}
    \centering
    \includegraphics[scale=0.21]{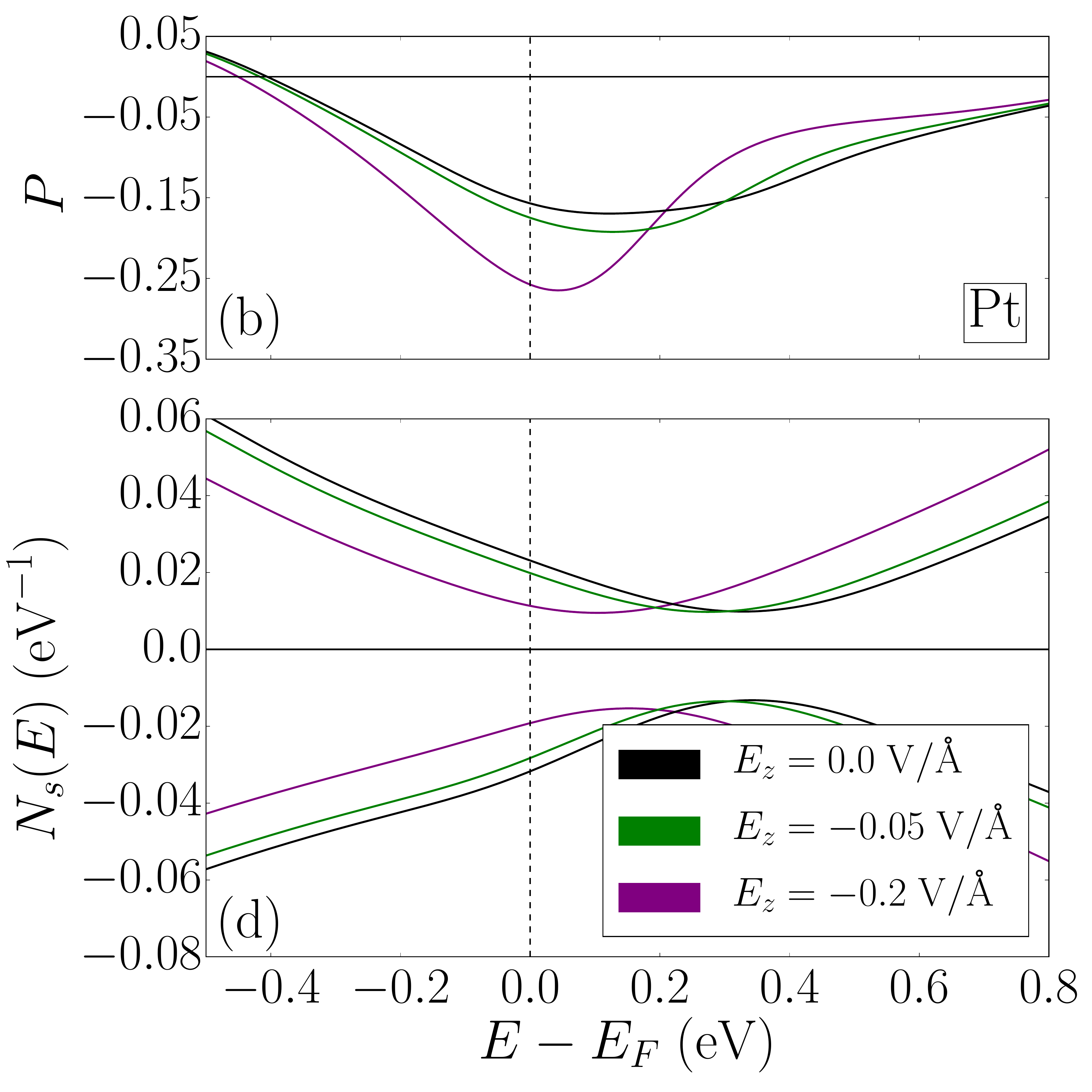}
   \end{subfigure}  
   \caption{[(a), (b)] Energy-resolved spin polarization of PDOS [Eq.~(\ref{polarization})] and [(c), (d)] PDOS in graphene in $\sysau$ [panels (a) and (c)] and $\syspt$ [panels (b) and (d)] with and without an applied electric field for 100\% coverage in the passivation layer. $N_\downarrow(E)$ is plotted with negative sign. The vertical dashed line denotes the Fermi level.
   }
   \label{PDOS}
\end{figure*}

Induced $\mathcal{S}$ in graphene is considerably larger in $\syspt$ compared to $\sysau$, owing to greater hybridization between the graphene and metal bands in former compared to latter. There are two reasons for this stronger hybridization. First, graphene is closer to $\Co$ in $\syspt$ than in $\sysau$, as seen in Fig.~\ref{system}(b), because $\Pt$ binds more strongly to $\Co$. 
Second, $\syspt$ features a band shown in the inset of Fig.~\ref{bandstructure}(b), which is absent in $\sysau$. Orbital analysis shows that this band is predominantly located on $\Pt$ atoms and the first underlying layer of $\Co$ atoms, i.e., $\syspt$ has a polarized surface resonant band near the Fermi level. This band naturally has larger overlap with graphene wave functions than the other, bulk-like, bands.

In the context of transport properties of proximity-magnetized graphene, only $N_s(E)$ and its polarization near the Fermi level, shown in Fig.~\ref{PDOS}, are important. Consistently with the band structure showing a largely intact Dirac cone, the shape of PDOS is similar to freestanding graphene. The avoided crossings with the bands of the metallic surface give rise to pronounced peaks and dips in graphene PDOS in both $\sysau$ and $\syspt$, but these peaks are not visible in Fig.~\ref{PDOS} due to the broadening used in the calculation of PDOS.

In $\syspt$ hybridization of the surface band seen in the inset of Fig.~\ref{bandstructure} with the graphene states results in a large $P(E_F)$. The avoided crossings in $\sysau$ are less pronounced, and their effect on the band structure does not extend to the Fermi level. As a result, $P(E_F)$ is smaller in $\sysau$.

Large $P(E_F)$ in $\syspt$ should be experimentally testable by spin-polarized scanning tunneling microscopy (STM) or scanning tunneling spectroscopy (STS) experiments~\cite{Hwang2016:SciRep,Choi2017:PRL,YangFang2013:PRL}. 

\subsection{Submonolayer coverage in the passivation layer}

In this Section we show that the doping level in graphene can be effectively controlled by changing the coverage in the passivation layer, which also has a strong effect on the spin polarization.

\begin{figure*}
\centering
  \begin{subfigure}
    \centering
    \includegraphics[scale=0.21]{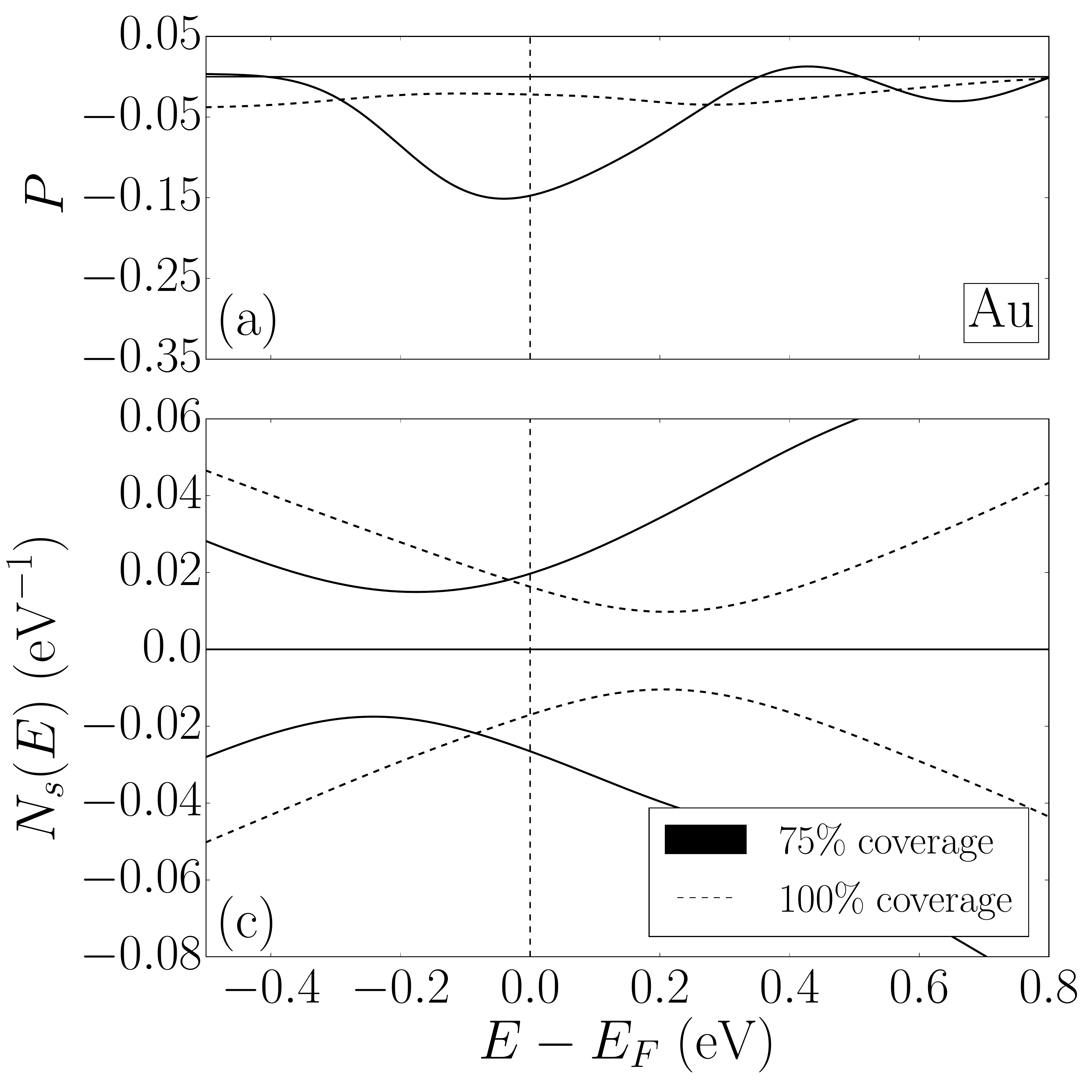}
   \end{subfigure}
\qquad
   \begin{subfigure}
    \centering
    \includegraphics[scale=0.21]{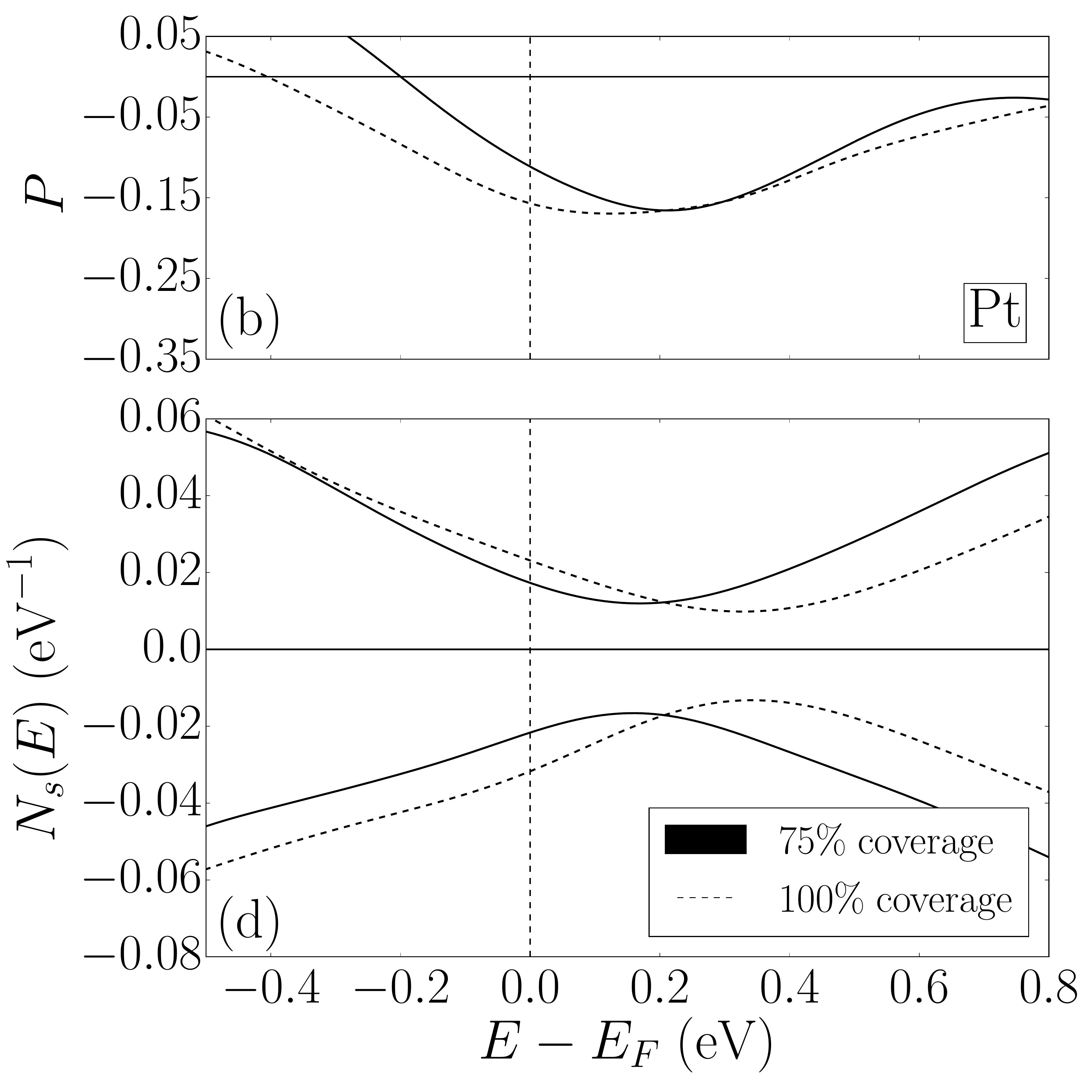}
   \end{subfigure}  
   \caption{ [(a), (b)] Energy-resolved spin polarization of PDOS [Eq.~(\ref{polarization})] and [(c), (d)] PDOS in graphene in $\sysau$ [panels (a) and (c)] and $\syspt$ [panels (b) and (d)]. 
   Solid (dashed) lines: $75\%$ ($100\%$) coverage in the passivation layer.
   }
   \label{PDOS_75}
\end{figure*}

Figure~\ref{PDOS_75} shows the graphene PDOS in $\sysau$ and $\syspt$ with $75\%$ coverage in the Au or Pt layer at zero external field. Based on the similarity of this PDOS to the DOS of freestanding graphene and the large distances between graphene and the metallic surface [$3.2\ \angstrom$ for $\sysau$ and $3.3\ \angstrom$ for $\syspt$; see Fig.~\ref{system}(d)], we again conclude that graphene is physisorbed. Smaller distance between graphene and the surface
with 75\% coverage can be attributed to stronger binding, similarly to graphene on $\Au$-passivated $\Ni$ surface~\cite{Kang2010:PRB}.

As seen in Fig.~\ref{PDOS_75}, the reduction in coverage has a significant effect on the proximization of graphene. First, the Dirac point is shifted from $0.22\ \electronvolt$ ($p$-type doping) to $-0.21\ \electronvolt$ ($n$-type doping) in $\sysau$ and from $0.33\ \electronvolt$ to $0.17\ \electronvolt$ in $\syspt$. This is similar to the shift observed in the Ni/Au/Gr system~\cite{Kang2010:PRB}. Second, $P(E_F)$ in $\sysau$ increases greatly for $75\%$ coverage. Orbital analysis shows that no surface bands are formed in the $\sysau$ system. Thus, the increased $P(E_F)$ is due to stronger hybridization of graphene and $\Co$ states due to the reduced distance. This explanation is consistent with the much smaller change in $P(E_F)$ in $\syspt$ where the graphene-$\Co$ distance does not decrease as much.

Therefore, the change in the coverage can drastically change the doping level and even switch it from $p$-type to $n$-type. Because doping in Ni/Au/Gr changes monotonically with coverage~\cite{Kang2010:PRB}, we expect there is a certain coverage between $75$ and $100\%$ in $\sysau$ (and, similarly, less than 75\% in $\syspt$) for which graphene is undoped. This choice of coverage could lead to greater sensitivity to the applied electric field.

\subsection{Gate control of spin polarization}

\begin{figure*}
\centering
\begin{subfigure}
\centering
  \includegraphics[scale=0.4]{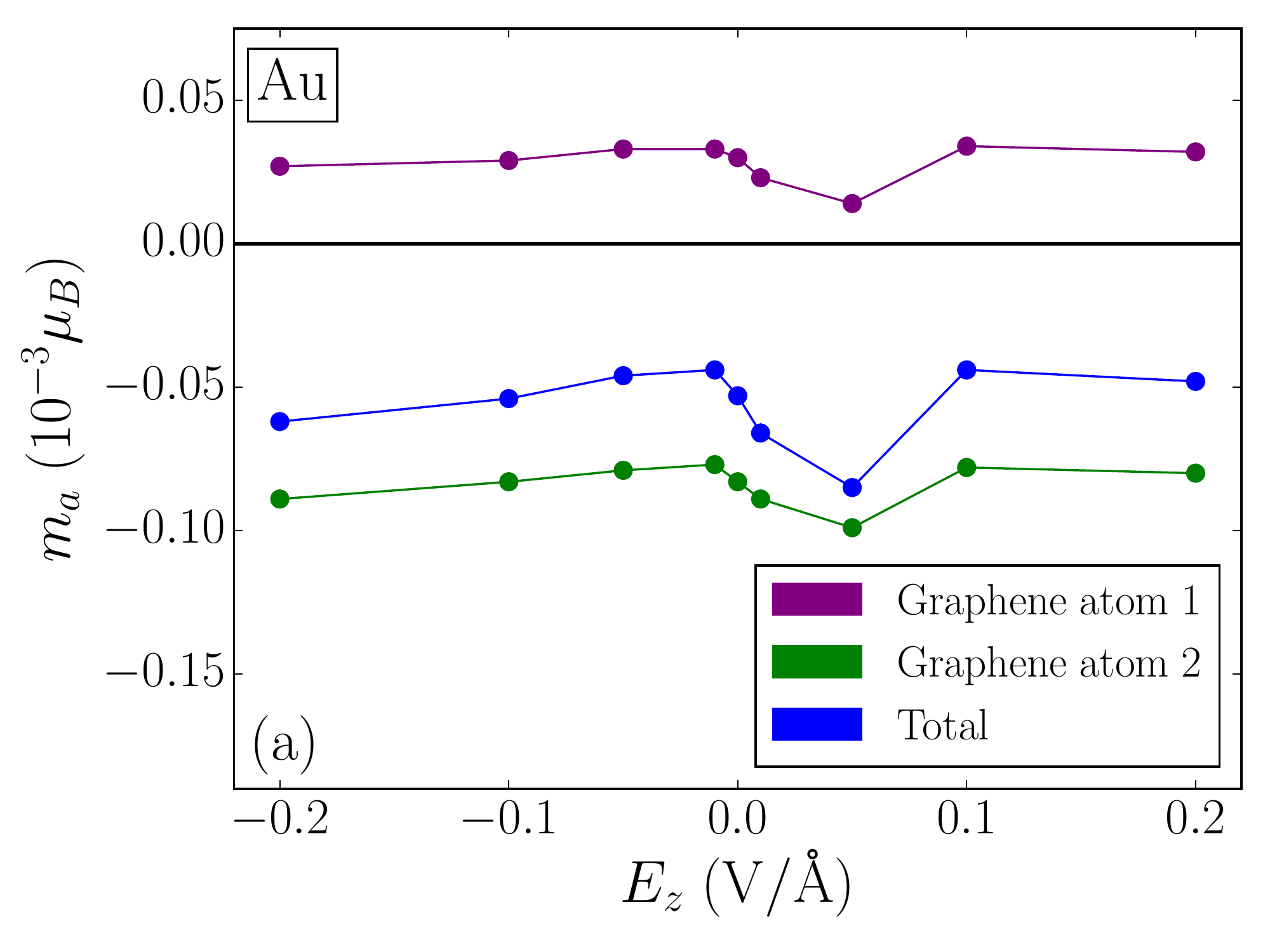}
\end{subfigure}
\quad
\begin{subfigure}
\centering
  \includegraphics[scale=0.4]{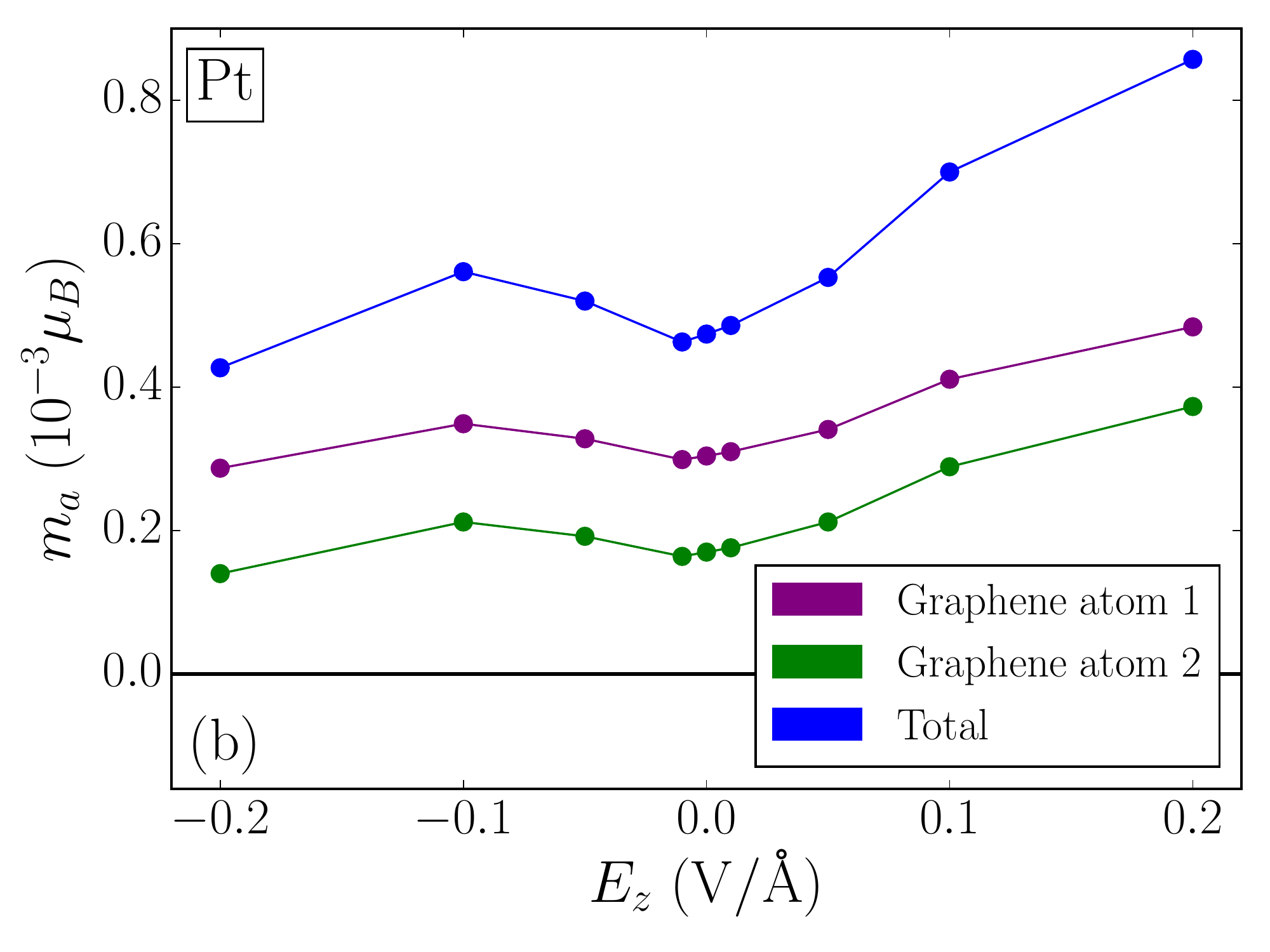}
\end{subfigure}
    \caption{
    Magnetic moments on the two inequivalent carbon atoms for (a) $\sysau$ and (b) $\syspt$ with full coverage in the passivation layer. The sites are labeled as indicated in Fig.~\ref{system}(a).
    }
    \label{atomic_polarizations}
\end{figure*}

The applied electric field induces charge transfer between graphene and the rest of the system, shifting the PDOS relative to the Fermi level, as seen in Figs.~\ref{PDOS}(c)-\ref{PDOS}(d) for $100\%$-covered systems. This shift can be traced through the position of the point where PDOS has a minimum. This point moves up (away from the Fermi level) at $E_z>0$ because the electronic states of graphene are raised in energy compared to the metal, and vice versa.

For the same magnitude of the electric field, total PDOS shifts more in $\sysau$ compared to $\syspt$. This is in agreement with the electrostatic model~\cite{Lazic2016:PRB}: doping in $\sysau$, and hence PDOS at $E_F$, is lower in $\sysau$ compared to $\syspt$.

On the other hand, Fig.~\ref{atomic_polarizations} shows the magnetic moments of the carbon atoms as a function of electric field in systems with 100\% coverage. The magnetic moments and their changes due to the electric field are greater in $\syspt$ than in $\sysau$. Furthermore, Figs.~\ref{PDOS}(a)-\ref{PDOS}(b) show $P(E)$ near the Fermi level as a function of the electric field. Similarly to the magnetic moments, $P(E)$ in $\syspt$ is more responsive to the field compared to $\sysau$ due to strong hybridization of the graphene states with the surface resonant band in $\syspt$.

Note that the change in $P(E)$ induced by the field is not limited to a simple shift in energy. This is because the bands of graphene and the metal surface are shifted relative to each other, which modifies their hybridization in a non-trivial way. 

In contrast to the Co/h-BN/Gr system where $P(E_F)$ changes sign in a large gate field~\cite{Lazic2016:PRB,Zollner2016:PRB}, it remains negative in the studied field range from $-0.2\ \volt/\angstrom$ to $0.2\ \volt/\angstrom$ in both $\sysau$ and $\syspt$.

\section{Conclusions and outlook}

We have studied the magnetic proximity effect in graphene that is physisorbed on a ferromagnetic $\Co$ slab passivated by a layer of $\Au$ or $\Pt$. The induced spin polarization can be tuned by the electric field, similarly to the previously studied systems where h-BN or graphene were used as passivation layers~\cite{Lazic2016:PRB, Zollner2016:PRB}.

The proximity-induced spin density in graphene has the same (opposite) sign on the two inequivalent carbon atoms in $\syspt$ ($\sysau$). The analysis of orbital contributions to the spin density shows that these patterns are produced by hybridization of both graphene orbitals with the $d$ bands of the metal.

For spintronic applications relying on transport~\cite{Tsymbal:2019}, the spin polarization of the bands near the Fermi level is important. We assume that the current can flow almost exclusively through the physisorbed graphene due to its separation from the underlying metal by the van der Waals gap. Note that this sets the limit on how long the contact region with the metal can be: the resistance along graphene sheet scales with the length of contact region L, while the tunneling resistance across vdW gap scales inversely with L. When L is under this limit the spin polarization $P(E_F)$ at the Fermi level of the projected DOS in graphene [see Eq.~(\ref{polarization})] can be used as a measure of the transport spin polarization in this system.

$P(E_F)$ is negative in both $\syspt$ and $\sysau$. Its magnitude is larger, and the response to the applied field stronger, in $\syspt$ compared to $\sysau$ when Co is fully covered by the passivation layer in $1\times1$ geometry. $P(E_F)$ in $\syspt$ is large enough to be resolved by spin-polarized STM or STS experiments.

Even in $\sysau$ the magnitude of $P(E_F)$ at zero field is large compared to Co/h-BN/Gr ~\cite{Lazic2016:PRB}, but the response to the electric field is only slightly stronger.

The large magnitude of $P(E_F)$ in $\syspt$ is due to a spin-down surface band just below the Fermi level. Strong response to the electric field in this system is mediated by the field-induced shift of this band relative to the Dirac point of graphene and is also facilitated by the lower doping of graphene compared to the systems with h-BN or graphene passivation layers. Because the response to the field is weaker in $\sysau$ where the doping level is even lower compared to $\syspt$, we conclude that the presence of a surface band near the Fermi level can serve as a design criterion for achieving strong response to the electric field in proximitized graphene.

In contrast with Co/hBN/Gr heterostructures~\cite{Lazic2016:PRB,Zollner2016:PRB,Asshoff2017:2DM}, graphene in fully covered $\sysau$ and $\syspt$ systems is $p$-doped. However, the doping level decreases, and even switches to $n$-type in $\sysau$, if the passivation layer coverage is decreased to 75\%. It is, therefore, expected that graphene should be undoped for a certain passivation layer coverage in both $\sysau$ and $\syspt$.

Further studies aiming at efficient gate-controlled spin polarization in graphene should concentrate on finding structures with nearly compensated graphene that feature surface bands near the Fermi level.

Elucidating magnetic proximity effects in graphene heterostructures with metallic ferromagnets remains an important issue as recent experiments on bias-dependent reversal of  magnetoresistance in vertical Co/Gr/NiFe spin valves support an overlooked role of graphene~\cite{Asshoff2017:2DM}. These experiments show that, in contrast to ideally lattice-matched single-crystalline ferromagnet/Gr structures required for effective spin filtering~\cite{Karpan2007:PRL}, van-der-Waals heterostructures without such lattice matching can be used in devices where proximitized graphene itself serves as a \emph{source} of spin-polarized carriers~\cite{Lazic2016:PRB,Asshoff2017:2DM}. While Co and NiFe were responsible for effective $n$- and $p$-doping of graphene
in these experiments~\cite{Asshoff2017:2DM}, our studies show that, even with a single ferromagnet, both $n$- and $p$-doping of graphene is possible, which could enable different approaches for designing bias-dependent 
effects~\cite{Asshoff2017:2DM,Ringer2018:APL,Zhu2018:PRB,Gurram2017:NC}. 
In particular, the gate-controlled modulation of spin polarization in Co/Pt/Gr based structures presented here would be desirable for transferring information in spin interconnects~\cite{Dery2011:APL,Zutic2019:MT}.

\begin{acknowledgments}
M.B.~and I.L.~acknowledge support from Croatian Science Foundation and the European Union through the European Regional Development Fund within the Competitiveness and Cohesion Operational Programme (Grant No.~KK.01.1.1.06).
K.D.B.~is supported by NSF through Grants DMR-1609776, DMR-1916275, and the Nebraska Materials Research Science and Engineering Center (MRSEC) Grant DMR-1420645. I.\v{Z}. is supported by the US ONR N000141712793, NSF ECCS-1810266, and the UB Center for Computational Research.
\end{acknowledgments}

\bibliography{references.bib}
\end{document}